\documentclass[preprint,showpacs,aps,amsmath]{revtex4}
\usepackage{graphicx}
\usepackage{bm}

\def\vec#1{{\bf #1}}
\def\half{ {1\over 2} }
\def\threehalves{ {3\over 2} }
\def\eps{\epsilon}
\def\avg#1{\left\langle #1 \right\rangle}
\def\LB{\left\lbrack}
\def\RB{\right\rbrack}
\def\LP{\left (}
\def\RP{\right )}
\def\pt{\partial}

\def\pxx#1{{\partial #1\over\partial x}}
\def\pyy#1{{\partial #1\over\partial y}}
\def\pss#1{{\partial #1\over\partial s}}
\def\ptt#1{{\partial #1\over\partial t}}
\def\pps#1{\partial #1/\partial s}
\def\grad{\nabla}
\def\div{\grad\cdot}
\def\cross{{\bf \times}}
\def\dpl{\grad_\parallel}
\def\ddpl{\grad_\parallel^2}
\def\dpp{\grad_\perp}
\def\ddpp{\grad_\perp^2}
\def\delsq{\grad^2}
\def\bdel{\vec b\cdot\grad}
\def\Bdel{\vec B\cdot\grad}
\def\udel{\vec u\cdot\grad}
\def\Jpl{J_\parallel}
\def\upl{u_\parallel}
\def\kappacv{{\cal K}}
\def\rs{\rho_s}
\def\ptb{\widetilde}
\def\phifl{\widetilde\phi}
\def\nfl{\widetilde n}
\def\Afl{\ptb A_\parallel}
\def\Jfl{\ptb J_\parallel}
\def\qqquad{\qquad\qquad}
\def\phig{\phi_G}
\def\vorg{\Omega_G}
\def\phigfl{\phifl_G}

\def\vorgfl{\ptb\Omega_G}

\def\ppyyk#1{{\pt^2 #1\over\pt y_k^2}}

\def\qeplfl{\widetilde q_e{}_\parallel}
\def\qeppfl{\widetilde q_e{}_\perp}

\def\tippfl{\widetilde T_i{}_\perp}

\def\nzfl{\widetilde n_z}

\def\uzfl{\widetilde u_z{}_\parallel}

\def\tzplfl{\widetilde T_z{}_\parallel}
\def\tzppfl{\widetilde T_z{}_\perp}
\def\qzplfl{\widetilde q_z{}_\parallel}
\def\qzppfl{\widetilde q_z{}_\perp}

\def\piplfl{\widetilde p_i{}_\parallel}
\def\pippfl{\widetilde p_i{}_\perp}
\def\pzplfl{\widetilde p_z{}_\parallel}
\def\pzppfl{\widetilde p_z{}_\perp}
\def\uexb{\vec u_E}
\def\wexb{\vec w_E}

\def\uex{u_E^{\;x}}
\def\wex{w_E^{\;x}}
\def\uedl{\uexb\!\cdot\!\grad}
\def\wedl{\wexb\!\cdot\!\grad}
\def\sumsp{\sum_z}
\def\drift{{c\over B^2}\vec B\cross}

\begin{document}
\preprint{}
\title{Nonlinear gyrofluid computation of edge localised ideal ballooning modes}
\author{Alexander Kendl}
\affiliation{Institut f\"ur Ionenphysik und Angewandte Physik, Association
  Euratom-\"OAW,  Universit\"at Innsbruck, A-6020 Innsbruck, Austria}
\author{Bruce D. Scott}
\affiliation{Max-Planck-Institut f\"ur Plasmaphysik, Euratom Association,
                D-85748 Garching, Germany}
\author{Tiago Ribeiro}
\affiliation{1) Max-Planck-Institut f\"ur Plasmaphysik, Euratom Association,
                D-85748 Garching, Germany\\
2) Instituto de Plasmas e Fus\~{a}o Nuclear, EURATOM/IST Association,
1049-001 Lisboa, Portugal}

\date{\today}

\begin{abstract}
Three dimensional electromagnetic gyrofluid simulations of the ideal
ballooning mode blowout scenario for tokamak edge localized modes (ELMs)
are presented.  Special emphasis is placed on energetic diagnosis,
examining changes in the growth rate in the linear, overshoot, and decay
phases.  The saturation process is energy transfer to self generated
edge turbulence which exhibits an ion temperature gradient (ITG) mode
structure.  Convergence in the decay phase is found only if the spectrum
reaches the ion gyroradius.  The equilibrium is a self consistent
background whose evolution is taken into account.  Approximately two
thirds of the total energy in the edge layer is liberated in the blowout.
Parameter dependence with respect to plasma pressure and the
ion gyroradius is studied.
Despite the violent nature of the short-lived process, the transition to
nonlinearity is very similar to that found in generic tokamak edge
turbulence.
\end{abstract}

\maketitle

\section{Introduction}

Edge localised modes (ELMs) are bursty, quasi-periodic expulsions of energy
and particles from the plasma edge in the high-confinement
state of toroidal magnetised plasmas \cite{Zohm96,Connor98a,Becoulet03}.
ELMs originate from the steep pressure gradient pedestal in the outer
closed field line region of a tokamak. 
A substantial release of magnetic and potential energy, caused by a rapid rise
of the rate of momentum transfer through the flux surfaces, leads to enhanced
transport and loss of heat and particles into the open scrape-off layer (SOL)
field line region. The filamented peak heat fluxes during ELMs on the bounding
divertor plates present serious restrictions on the performance of future
fusion experiments like ITER \cite{ITER} and necessitate the development of
suitable control techniques \cite{Lang03,Evans06,Hand08}.

ELMs generically have been observed in all divertor tokamaks since the initial
discovery of the high-confinement ``H-mode'' state in the early 1980s
\cite{Wagner82}, while characteristics like frequency and intensity are found
to vary widely depending on experimental conditions \cite{Kamiya07}. 
Stellarator experiments also are reported to show similar bursty
quasi-periodic edge activity in the presence of edge transport barriers
\cite{Wagner06}.  Indication that equilibrium drifts are part of the
process has been found by experimental diagnosis of an asymmetry of
the energy and particle fluxes on each of the divertor plates and
evidence that this is sensitive to the directions of the toroidal
current and magnetic field
\cite{Eich03,Eich07,Eich09}.

Phenomenologically, edge localised mode events in tokamaks have been
compared to solar flare eruptions \cite{Wilson04,Fundamenski07}, and, in the
cataclysmic variability of observed radiation emanating from the plasma, they
may seem to bear also some remote resemblance to the outbursts of
pressure-driven dwarf novae \cite{Bath85}, although the specific instability
mechanisms behind these phenomena are clearly of a substantially different
nature.  

On the other hand, the nature of ELMs has also not yet been completely
clarified, in the sense that there is no first-principles based predictive
theory available. A number of characteristically different variations
of ELMs have been observed in experiments (e.g. Type I, II or III ELMs) which
may actually be caused by different instability mechanisms \cite{Connor98b}. 
``Type III'' ELM observations show a remarkable similarity to global bursts 
found in computations of
drift-wave turbulence, where the turbulence generated flows and fluxes are
closely tied to the self-consistent equilibrium evolution in the plasma edge
\cite{Scott06,Scott07}. 

``Type I'' ELMs in the experiment specifically show stronger magnetic activity
during the burst and are usually associated with the onset of a 
magneto-hydrodynamic (MHD)
ideal ballooning mode (IBM) instability when a threshold 
pressure gradient is reached in the steep edge pedestal of an H-mode
plasma \cite{Connor98b}. 
The possible role of an additional current driven instability
(``peeling-ballooning mode'') for ``Type I'' ELMs 
has also received recent interest \cite{Snyder02,Wilson06,Snyder07}.

The ideal ballooning instability with nonlinear phases or aftermath has
provided a paradigm for recent and current study of the ELM phenomenon
theoretically.  Explosive MHD instability and associated critical
phenomena have been advanced analytically \cite{Wilson04}.
Early Braginskii fluid simulations were given in support of this
\cite{Snyder05}.  Large, nonlinear MHD codes have been studying this in
more detail, resolving several numerical problems and paying specific
attention to the details of the magnetic geometry,
with mixed results on the phenomenology of the instability beyond its
linear phase
\cite{Brennan06,Strauss06,Pankin07,Mizuguchi07,Huysmans07}.
Whether or not the ideal ballooning instability or its
peeling/ballooning variant is actually responsible for ELMs occuring in
experiments (here noting the absence of a demonstrably well-resolved 
L-to-H mode transition in computations from first principles),
the nonlinear phase of this ``ideal MHD blowout'' or ``IBM blowout''
phenomenon is of physical interest.  The ability of the instability to
destroy the edge pedestal layer is undisputed, and the term ``blowout''
is germane.  Typical physical parameters lead to a situation wherein the
ideal MHD interchange growth rate is comparable to the parallel Alfv\'en
transit time, one and the same with the basic ideal ballooning criterion.
This is fast with respect to MHD but not to microturbulence.  Provided
they can simultaneously treat global MHD (i.e., the spectrum covers both
global and ion gyroradius scales), edge-turbulence computations can
also address the phenomenon and are even in a position to treat parts of
it which fall outside the paradigm of the MHD model.

Here we present nonlinear gyrofluid computations of IBM blowout events
localised to the edge/SOL region of a tokamak plasma, resolved to below
the ion gyroradius $\rho_i$.
As in the large MHD models the self-consistent
evolution of the equilibrium is not only included but is also an
integral part of the dependent variables.  An electromagnetic 6-moment
gyro-fluid model (``GEMR'') for both electrons and ions is used,
including an energy-conserving treatment of finite ion Larmor radius
effects and higher-moment terms which occur in the toroidal drift
\cite{Scott05}.
The energetic consistency is a key feature since the reaction of parts
of the system such as flows and currents which have low energy content
can play a central role in the indirect nature of nonlinear dynamics by
serving as transfer channels.  Although the gyrofluid model has a
different formulation of nonlinear polarisation (the generalisation of
``vorticity'' to a two-fluid setting) than a fluid or MHD model, full
correspondence in the regime of validity of the latter models has been
shown \cite{Scott07b}.  In this sense the model is a superset of global
reduced MHD and electromagnetic microturbulence.  The global dynamics is
in the reduced MHD regime \cite{Strauss76} due to low absolute edge beta
values ($\beta=8\pi p/B^2\sim 10^{-3}$) but the small scales depend on
treating the ion gyroradius to arbitrary order.  The model does not
assume instabilities to occur at any particular scale, but when they
occur the resulting spread of the spectrum is found to reach $\rho_i$
within a few eddy turnover times of the onset of nonlinearity in every
case computed.

The work reported herein addressed the following points.  The
relationship of the MHD dynamics to the ideal threshold is much
discussed but in the gyrofluid model several mechanisms of
microinstability are also present.  The threshold issue is obscured by
the existence of ion temperature gradient (ITG) drift instabilities at
all beta values, with a smooth transition between them.  Nevertheless,
varying character in the saturation phases is observed at different
values.  We also address the issue of possible ``rho-star'' dependence
(here the local rho-star is given by $\rho_s/L_T$ where $\rho_s$ is
$\rho_i$ evaluated at $T_i=T_e$ and $L_T$ is the scale length of the
$T_e$ profile, both evaluated at the mid-pedestal location in minor
radius) which would indicate a role for diamagnetic drifts.  The
gradient is sufficiently steep that $\rho_s/L_T > L_T/qR$ so that none
of these effects can be ordered small.  We examine the necessary
resolution to obtain converged cases, which gives the spectral range
actively involved in nonlinear saturation.  The main instability at
experimentally relevant beta is the ideal ballooning instability but the
main saturation process is energy exchange with broadband
electromagnetic drift wave/ITG turbulence which the instability itself
generates.  Due mostly to this, the results turn out to be outside
of both the magneto-hydrodynamic (MHD) and collisional Braginskii
paradigms, on which most previous approaches have been based.
On the other hand, the GEMR model still uses delta-f equations, so the
actual profile phenomenology in the SOL region is not well represented,
and as in other work with GEMR we concentrate on the properties of the
nonlinear dynamics \cite{Zweben09}.

Following sections give the details of the GEMR model, a discussion of
the physical difference between ITG and MHD dynamics and their relative
roles, the methods used to pre-set the axisymmetric equilibrium state so
that the computations are done in the absence of axisymmetric
oscillations whose decay times can be slower than the IBM rise time, the
computations of the blowout phase itself, and the possibility and
limitations of quantitative comparison to experiment.

\section{The gyrofluid model with radially dependent axisymmetric geometry}

The gyrofluid model used herein is given in Ref.\ \cite{Scott05}. It is
based on the original derivations given in Refs.\
\cite{Dorland93,Beer96}, with corrections given as motivated by free
energy conservation.  An alternative derivation using the conservation
laws of the underlying gyrokinetic model is given in Ref.\
\cite{gem2}. The equations are normalised to spatial scale $a$ and time
scale $a/c_s$ where $a$ is the minor radius and $c_s$ is the reference
sound speed given by $c_s^2=T_e/M_i$, that is, in terms of the electron
temperature (in energy units) and ion mass. The main parameters are the
drift parameter, electron dynamical beta, and electron collisionality,
respectively given by
\begin{equation}
\rho_* = {\rs\over a} \qqquad
\beta_e = {4\pi p_e\over B^2} \qqquad
\nu_e = {a\over c_s\tau_e} 
\end{equation}
where $c$ and $e$ are the speed of light and fundamental charge, $p_e$
and $\tau_e$ are the pressure and Braginskii collision time
\cite{Braginskii} of the electrons, and $\rs=c_sM_i c/eB$ 
is the drift scale.  Note that if $T_i=T_e$ then $\rs=\rho_i$.  The
reference for $M_i$ is the deuterium mass $M_D$.  For the electrons the
physical value of the mass $m_e/M_D = 1/3670$ is always used.  With
global geometry there is no single magnetic field line connection
length, but one with a profile with $2\pi qR_0$ a function of the minor
radius coordinate, where $R_0$ is a constant giving the reference value
of the major radius.

\subsection{Species constants and gyrofluid moment equations}

Each species $z$ has its own set of gyrofluid moment variable equations,
one each for density $\nzfl$, parallel velocity $\uzfl$, parallel and
perpendicular temperature $\tzplfl$ and $\tzppfl$, and parallel and
perpendicular energy components of the parallel heat flux $\qzplfl$ and
$\qzppfl$.  They are coupled to the electrostatic and parallel magnetic
potentials $\phifl$ and $\Afl$ through self consistent field equations.
The tilde denotes a dependent variable to distinguish from
constant parameters.  The correspondence of these equations to the
Braginskii fluid model is given in Ref.\ \cite{Scott07b}.  Each species
is characterised by a background charge density, temperature/charge
ratio, and mass/charge ratio, given by the normalised parameters
\begin{equation}
a_z = {Zn_z\over n_e}  \qqquad
\tau_z = {T_z\over ZT_e}  \qqquad
\mu_z = {M_z\over ZM_i}.
\end{equation}
For electrons $a_z=\tau_z=-1$ and $\mu_z=-m_e/M_i$.  For the main ions
$a_z=\mu_z=1$ while $\tau_z=\tau_i$ is kept as a parameter.  For the
trace ions $a_z=0$ is always taken.  For hot trace ions $\tau_z$ is
large while $\mu_z$ is still moderate.  The species gyroradius $\rho_z$
is given by $(\rho_z/\rho_s)^2 = \mu_z\tau_z$, which is always small for
electrons, close to unity for main ions, and can be moderate or large
for trace ions.  For this study a simplified geometry is taken with
$B=1$ except in curvature terms, so magnetic pumping and shaping effects
\cite{Kendl06} are neglected.  
Since dynamics at the electron gyroradius scale are neglected the
gyroaveraging effects on $\Afl$ are not treated.  

The moment equations are given by
\begin{eqnarray}
& &\ptt{\nzfl}+\uedl\nzfl+\wedl\tzppfl + \dpl\uzfl 
\nonumber\\ & & \qqquad{}
	= \kappacv\LP\phigfl+{\vorgfl\over 2}
	+\tau_z{\pzplfl+\pzppfl\over 2}\RP
\end{eqnarray}
\begin{eqnarray}
& &\ptt{}\LP\beta_e\Afl+\mu_z\uzfl\RP
	+ \mu_z\uedl\uzfl+\mu_z\wedl\qzppfl 
\nonumber\\ & & \qqquad{}
	+ \dpl\LP \phigfl+\tau_z\pzplfl\RP
\nonumber\\ & & \qqquad{}
	= \mu_z\tau_z\kappacv\LP{4\uzfl+2\qzplfl+\qzppfl\over 2}\RP
	- R_{ei}
\end{eqnarray}
\begin{eqnarray}
& &\half\ptt{\tzplfl}+\half\uedl\tzplfl + \dpl(\uzfl+\qzplfl)
\nonumber\\ & & \qqquad{}
	= \kappacv\LP{\phigfl+\tau_z\pzplfl+2\tau_z\tzplfl\over 2}\RP
	- S_\Delta
\end{eqnarray}
\begin{eqnarray}
& &\ptt{\tzppfl}+\uedl\tzppfl+\wedl(\nzfl+2\tzppfl) + \dpl\qzppfl 
\nonumber\\ & & \qqquad{}
	= \kappacv\LP{\phigfl+\vorgfl+\tau_z\pzppfl\over 2}
	+3{\vorgfl+\tau_z\tzppfl\over 2}\RP
\nonumber\\ & & \qqquad{}
	+ S_\Delta
\end{eqnarray}
\begin{eqnarray}
& &\mu_z\ptt{\qzplfl} + \mu_z a_L(\qzplfl)
	+ \mu_z\uedl\qzplfl
	+ \threehalves\tau_z\dpl\tzplfl
\nonumber\\ & & \qqquad{}
	= \mu_z\tau_z\kappacv\LP{3\uzfl+8\qzplfl\over 2}\RP
	- K_\parallel - K_\Delta
\end{eqnarray}
\begin{eqnarray}
& &\mu_z\ptt{\qzppfl} + \mu_z a_L(\qzppfl)
	+ \mu_z\uedl\qzppfl
\nonumber\\ & & \qqquad{}
        + \mu_z\wedl(\uzfl+2\qzppfl)
	+ \dpl\LP \vorgfl+\tau_z\tzppfl\RP
\nonumber\\ & & \qqquad{}
	= \mu_z\tau_z\kappacv\LP{\uzfl+6\qzppfl\over 2}\RP
	- K_\perp +  K_\Delta
\end{eqnarray}
The linearised pressures are given by
\begin{equation}
\pzplfl = \nzfl+\tzplfl  \qqquad 
\pzppfl = \nzfl+\tzppfl
\end{equation}
Collisional dissipation is controlled by a collision parameter 
$\nu_z$ analogous to $\nu_e$ and a set of numerical constants
for each species,
\begin{equation}
\alpha_z  \qqquad \kappa_z  \qqquad \pi_z
\end{equation}
which are the thermal force, thermal conductivity, and viscosity
coefficients involved in parallel dissipation, with values given by the
Braginskii model \cite{Braginskii}, though that regime is never reached
in core turbulence.  For electrons these coefficients are $0.71$ and
$3.2$ and $0.73$, and for ions they are $0$ and $3.9$ and
$0.96$, respectively.
The resistive dissipation is given by
\begin{equation}
R_{ei} = {m_e\over M_i}\nu_e\LB 0.51\Jfl + {0.71\over 3.2}
	\LP\qeplfl+\qeppfl+0.71\Jfl\RP\RB
\end{equation}
with parallel current given by
\begin{equation}
\Jpl = \sumsp a_z\uzfl
\end{equation}
noting that it is the electron heat fluxes that enter $R_{ei}$ and as
with $\beta_e\Afl$ the $R_{ei}$ term enters the parallel motion in the
same way for every species.
The anisotropy dissipation is given by
\begin{equation}
S_\Delta = {\nu_z\over 3\pi_z} \LP\tzplfl-\tzppfl\RP
\end{equation}
The thermal conduction components are given by
\begin{equation}
K_\parallel = \mu_z\tau_z\nu_z{5/2\over\kappa_z}\LP\qzplfl+0.6\alpha_z\Jfl\RP
\end{equation}
\begin{equation}
K_\perp = \mu_z\tau_z\nu_z{5/2\over\kappa_z}\LP\qzppfl+0.4\alpha_z\Jfl\RP
\end{equation}
\begin{equation}
K_\Delta = 1.28\mu_z\tau_z\nu_z{5/2\over\kappa_z}\LP\qzplfl-1.5\qzppfl\RP
\end{equation}
The Landau damping effects are modeled by
\begin{equation}
a_L = {V_z\over qR_0}\LP 1-0.125q^2R_0^2\ddpl\RP
\end{equation}
with thermal velocity $V_z$ given by $V_z^2=\tau_z/\mu_z$.

The gyroaveraging is done through Pad\'e approximants to operators which
would act in Fourier space,
\begin{equation}
\Gamma_1 = \LP 1 - \half\rho_z^2\ddpp\RP^{-1} 
\quad \mbox{and} \quad
\Gamma_2 = \rho_z^2{\pt\Gamma_1\over\pt\rho_z^2}
\end{equation}
and then the gyroaveraged potentials are
\begin{equation}
\phig = \Gamma_1\phi  
\qquad \mbox{and} \qquad 
\quad \vorg = \Gamma_2\phi
\end{equation}
noting that they are species dependent.  The main ExB advection and the
FLR correction pieces are given by
\begin{equation}
\uedl f = [\phig,f] - \nu_\perp\dpp^4 f + \nu_\parallel\ddpl f
\end{equation}
\begin{equation}
\wedl f = [\vorgfl,f]
\end{equation}
where the $[,]$ symbols denote the nonlinear bracket defined below,
and $\nu_\perp,\nu_\parallel$ denote the artificial dissipation
coefficients.  The parallel derivative is given by
\begin{equation}
\dpl f = {1\over qR_0}\pss{f} - \beta_e[\Afl,f]
\end{equation}
The curvature operator is given by
\begin{equation}
\kappacv(f) = - [\log B^2,f]
\end{equation}

\subsection{Field equations for polarisation and induction}

The species are coupled through two equations which give the self
consistent response of the field potentials to the evolution of the
gyrofluid moment variables.  The electrostatic potential is governed by
quasineutrality, which sets the total space charge density to zero.  The
density for each species is given by a part due to the gyrocenters and
another due to polarisation, which is what sets $\phi$.  This
polarisation equation is given by
\begin{equation}
{1-\Gamma_0\over\tau_i}\phi = \sumsp a_z\LB\Gamma_1\nzfl + \Gamma_2\tzppfl\RB
\end{equation}
where 
\begin{equation}
\Gamma_0 = \LP 1 - \rho_i^2\ddpp\RP^{-1}
\end{equation}
gives the gyroscreening of the main ions.  In this version of the model
the electron and trace ion contributions to gyroscreening
are neglected, due to small $m_e$ and zero $a_z$, respectively.  On
the right hand side the electrons and main ions enter with oppositely
signed $a_z$ values and again here the trace ions do not enter.  This
establishes them as a trace species.

In a similar way the trace ions are left out of the induction equation
due to the zero $a_z$, so that
\begin{equation}
-\rho_*^2\ddpp\Afl = \Jfl 
\qquad \mbox{and} \qquad
\Jfl = \sumsp a_z\uzfl
\end{equation}
determines $\Afl$ noting the way that the normalisation scales it with
respect to $\rho_*$.

More detail, including energy conservation and the relationship of these
equations to it, is given in Refs.\ \cite{Scott05,gem2}.

\subsection{Representation of the self consistent magnetic geometry}

In these expressions the operators $\ddpp$ and $\pps{}$ and the
nonlinear brackets are determined by the representation of the geometry
that is used.

The magnetic geometry is a simplified representation of an axisymmetric
magnetic field using field aligned Hamada coordinates.   The field
representation is
\begin{equation}
\vec B = I\grad\varphi + \grad\Psi\cross\grad\varphi
\end{equation}
where $\varphi$ is the physical toroidal angle about the symmetry axis,
$\Psi$ is the magnetic flux function, and $I=I(\Psi)$ is given by
the constraint of MHD equilibrium.
Without loss of
generality one may define magnetic flux coordinates $\{V,\theta,\zeta\}$
with the following
properties,
\begin{equation}
B^V = 0  \qqquad B^\theta = \chi_{,V}  \qqquad B^\zeta = \psi_{,V}
\end{equation}
where $V=V(\Psi)$ is the volume enclosed by the surface with flux $\Psi$,
then
$\chi=\chi(V)$ and $\psi=\psi(V)$ are two further flux functions found,
and the subscript denotes the partial derivative with respect to $V$.
The ratio $q=d\psi/d\chi$ is another flux function which gives the pitch
of the magnetic field.  The angles are cyclic on $[0,1]$ which
determines the coordinate Jacobian to be unity.  Hence the metric tensor
$g_{ij}$ satisfies $\det g_{ij}\equiv g = 1$.  The function $\chi$ is
found by the constraint that $\theta$ is cyclic on the unit interval
\begin{equation}
\chi_{,V}^{-1} = \oint {d\eta\over\Bdel\eta} 
\quad \mbox{and} \quad
\theta = \chi_{,V} \int {d\eta\over\Bdel\eta}
\end{equation}
where $\eta$ is a simple cyclic coordinate on the flux surface in the
poloidal plane (e.g., path length on the curve, or the physical angle
position about
the magnetic axis 
with respect to any branch cut).  Then $q$ is found 
\begin{equation}
q = {I\over 2\pi\chi_{,V}}\avg{{1\over R^2}}
\end{equation}
where the angle brackets give
the flux surface average defined as
\begin{equation}
\avg{{1\over R^2}} = \oint{d\theta\over R^2} 
        = \chi_{,V}\oint{d\eta\over R^2\Bdel\eta}
\end{equation}
Now, $\psi_{,V}=q\chi_{,V}$ is also defined and the toroidal angle
coordinate is given by
\begin{equation}
\zeta = {\varphi\over 2\pi} 
        + {I\over 2\pi\chi_{,V}}\int d\theta
	\LP \avg{{1\over R^2}} - {1\over R^2}\RP
\end{equation}
which completes the prescription.

The field aligning is a one-to-one and onto 
coordinate transform from $\{V,\theta,\zeta\}$ to $\{x,y,s\}$ given by
\begin{equation}
x = V/a^3  \qqquad y_k = q\theta - \zeta - \alpha_k(x) \qqquad s = \theta
\label{eqcoords}
\end{equation}
where $a$ is a reference minor radius and $\alpha_k$ is an arbitrary
function of $x$ which is chosen to make the off diagonal perpendicular metric
element $g^{xy}_k$ vanish at a particular location.  This is called the
shifted metric procedure \cite{Scott01}.  
The point is that at any position in $s$
where perpendicular drifts or $\ddpp$ is evaluated, the coordinate
elements are rectangular, and the information contained in magnetic
shear enters only in derivatives in the parallel coordinate.  The
magnetic field components now satisfy
\begin{equation}
B^x = 0  \qqquad B^y_k=0  \qqquad B^s = \chi_{,V}
\end{equation}
at all locations in space for any choice of $\alpha_k$.  Hence although
the perpendicular coordinate elements are rectangular only for one particular
location in $s$ the field aligning is exactly satisfied everywhere.

The coordinate metric elements $g^{ij}=\grad x^i\cdot\grad x^j$
are found in the $\{V,\theta,\zeta\}$ representation.
Then the grid locations $s=s_k$ are all given their own
members of the family of these field aligned coordinates via
\begin{equation}
\alpha_k(V) = q\,s_k + \left.
        \int_0^V dV'\,{q\,g^{V\theta}-g^{V\zeta}\over g^{VV}}
	\right\vert_{\theta=s_k}
\label{eqalphak}
\end{equation}
Then we have
\begin{eqnarray}
& g^{xx} = & a^{-6}g^{VV} \\
& g^{xy}_k = & 0 \\
& g^{yy}_k = & q^2\,g^{\theta\theta}-2q\,g^{\theta\zeta}+g^{\zeta\zeta}\\
& g^{xs} = & a^{-3}g^{V\theta}\\
& g^{ys}_k = & q\,g^{\theta\theta}-g^{\theta\zeta} \\
& g^{ss} = & g^{\theta\theta}
\end{eqnarray}
at $s=s_k$.
The drift tensor elements are given by
\begin{equation}
\vec F = \eps\cdot\vec B \qquad \mbox{with} \qquad F_{xy} = \chi_{,V}
\end{equation}
that is, $F_{xy}$ is the only independent, nonvanishing component.
We have
\begin{equation}
\grad f\cdot\drift\grad\phi \equiv 
    \grad\phi\cdot{c\vec F\over B^2}\cdot\grad f
\end{equation}
which defines the bracket $[\phi,f]$ for any scalar field quantities
$\phi$ and $f$.  With $\phi$ the electrostatic potential this gives the
ExB advection term, for electrons for which FLR corrections are
neglected by taking $\rho_e\to 0$.  For ions this is generalised by the
FLR corrections as given above.

For local equations as those used here (nonlinearities kept only
in advection and magnetic flutter) the geometry model must be modified
to retain free energy conservation by the equations.  In particular, any
function of $x$ such as a radially varying normalised parameter (such as
the temperature/charge ration $\tau_z$) multiplying any of the curvature
terms will cause them to fail to conserve free energy.  The
compressibility is already split away from advection and placed into the
curvature terms, so the brackets must retain the properties of
incompressible advection.  This means that the model for $\vec F$ which
is used must satisfy
\begin{equation}
\vec F\to\vec F_0 \qquad \mbox{with} \qquad \div {c\vec F_0\over B^2} = 0
\end{equation}
and to avoid any confusion the curvature terms are written with $\log
B^2$, which can be given any spatial dependence.  Correspondingly, the
MHD version of the 
continuity equation (neglecting diamagnetic fluxes)
has already been manipulated according to
\begin{equation}
\ptt{n}+\div n\vec u = 0
\end{equation}
\begin{equation}
\LP\ptt{}+\udel\RP\log n + \div \vec u = 0
\end{equation}
and then terms such as $\upl\bdel$ are neglected due to the ordering.
The result is the same as the local continuity equation
\begin{equation}
\LP\ptt{}+\udel\RP\nfl + n_0\div \vec u = 0
\end{equation}
where $n_0$ is a normalising constant,
if we identify the dependent variable $\nfl/n_0$ with $\log n$.  It is
important to note that with the self consistent background being evolved
this also includes the profile: $\grad\log n\to\grad\nfl$.

The simplified geometry is now defined by its operators.  The 
Laplacian and gyroaverging operators
neglect $\pps{}$.  The form
in the Ampere's law (which does not involve the gyroradius)
is given by
\begin{equation}
\delsq = \pxx{}g^{xx}\pxx{} + g^{yy}_k\ppyyk{}
\end{equation}
The form in the gyroaverging operations and polarisation is given by
\begin{equation}
\ddpp =  \pxx{}{g^{xx}\over B^2}\pxx{} + {g^{yy}_k\over B^2}\ppyyk{}
\end{equation}
The brackets are defined by
\begin{equation}
[f,g]=q[f,g]_{xy}+[f,g]_{xs}-q_{,x}(s-s_k)[f,g]_{ys}
\end{equation}
for any two scalar fields $\{f,g\}$
with each bracket piece defined by
\begin{equation}
[f,g]_{ij} = \LP{\partial f\over\partial x^i}{\partial g\over\partial x^j}
        - {\partial g\over\partial x^i}{\partial f\over\partial x^j}\RP
\end{equation}
with the third piece not used since these are always evaluated at
$s=s_k$.

\section{Mode character: ITG vs. MHD}

Nonlinear computations on ideal or peeling ballooning mode ELM scenarios in
tokamaks have previously been based on magnetohydrodynamic (MHD) models
\cite{Pankin07,Mizuguchi07,Snyder07,Huysmans07}, and the respective codes had
largely been tailored towards mode structures expected by linear analysis. 
Restriction to single-fluid MHD equations including two-fluid correction terms
allows computations of large to mesoscale dynamics
in realistically shaped 2D tokamak equilibria including
X-point geometry crossing the separatrix, due to the lack of a need to
use field-aligned coordinates.
Recent MHD simulations are able to reproduce spatio-temporal mode structures,
divertor footprints and other characteristics with good agreement to experimental
diagnostics, in particular for the growth and immediate blowout phase of the
finger-like instability \cite{Huysmans09}. 

However, as soon as the dynamics becomes nonlinear, the spectrum
broadens to include scales normally associated with drift wave dynamics.
This turbulence is basically of a drift-Alfv\'en type with strong edge ITG
character (see Ref.\ \cite{Scott02} for the signatures of the various
mode structure types), made more powerful than otherwise by the
energetic access to the long-wavelength MHD component.  The ITG
character results from the steeper logarithmic gradients in both
temperature profiles, and the lack of involvement of ion temperature
fluctuations in the parallel nonadiabatic response of the
electrons.  Although edge turbulence of this type (cf.\ \cite{Scott07})
is strongly suppressed in the H-mode phase and is not initially involved
in the instability phase, the experience of edge turbulence becomes
relevant as these physical components become involved in all
the nonlinear phases of the blowout.
The saturation and aftermath of the blowout, which in an experiment
carries a
large part of the actual transport losses that lead to a degradation of
the pedestal, should be expected to involve physics not contained in the
MHD model.

Ion temperature gradient (ITG) driven modes and ideal ballooning modes are
both essentially caused by the gradient and curvature driven interchange
instability and show similar character in their initial linear growth phases.
Instability is achieved above their respective critical gradients, determined
by the ratio $\eta_i = L_n / L_{Ti} > \eta_c$ between density gradient to ion
temperature gradient scale lengths for ITG, and by the ideal MHD
ballooning parameter $\alpha_M = q R \nabla \beta > \hat s$ 
for IBM, where $\hat s$ is the magnetic shear parameter and $\eta_c$ is
a critical threshold which depends on beta, collisionality, and
toroidicity. 
The principal difference between these modes
is the relative role of the parallel Alfv\'en responses which tend
towards enforcement of an adiabatic response in the electrons
(parallel force balance, with the
electron pressure gradient).  This is nonexistent in an MHD model and
subdominant generally for an MHD instabililty. But it constrains the
electrons if the $\alpha_M$ parameter is below threshold, still allowing
the ITG instability because the adiabatic response does not involve the
ions.  The ITG instability exists at all $\alpha_M$ values, and
furthermore for the edge situation the ITG instability and mode
structure transitions smoothly to a drift wave one
in a nonlinear setting for $\eta_i\sim 1$.  Hence there is actually no
finite threshold in values of $\eta_i$ or $\alpha_M$ for the nonlinear
edge situation.   Therefore, the
existence of a threshold in the experiment is the same as the mechanism
which maintains the H-mode, which is not yet well understood.
It follows that a gyrofluid model computation of an ELM scenario
with enough
resolution to allow for ion-gyroradius based dynamics has to face the
lack of a simple linear threshold in the parameter space.

The transition from initial (micro-)instabilities to generic edge turbulence
was studied in detail in Ref.~\cite{Scott05b}. As the most unstable linear
modes crystallise out of an initial random bath of small-amplitude
perturbations, the linear growth rate rises and becomes steady. The maximum
value of the instantaneous growth rate of total fluctuation free
energy $E$ given by  
$\Gamma(t) = (1/2E) (\partial E/\partial t)$
may be taken as the maximal linear growth rate. The curve of $\Gamma(t)$ then
falls very sharply to zero (over about $10 L_{\perp}/c_s$) as
saturation occurs. There is some  structural adjustment over the next few $100
L_{\perp}/c_s$ as the spectrum fills out, and then the turbulence is fully
developed. But over the adjustment phase the value of $\Gamma$ is well below
its previous maximum.  We will use the same diagnostic herein, except
that the total ion ExB heat flux $Q_i$ averaged over the computational
domain is used instead of $E$ because
most of $E$ is represented by the self consistent profiles while $Q_i$
is entirely due to fluctuations.

\begin{figure}
\includegraphics[width=5.0cm]{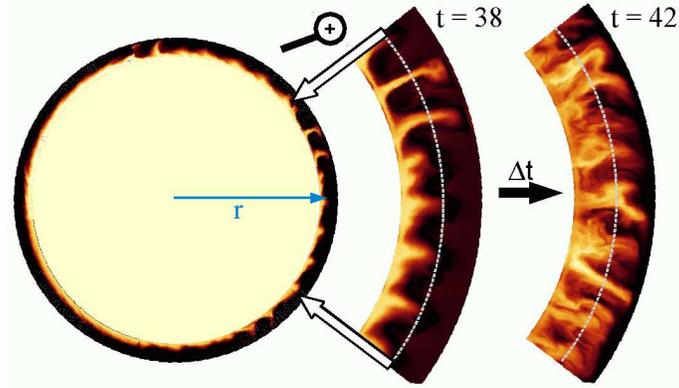}
\caption{\label{f:visual} Visualisation of the tokamak ELM burst: the edge
  region of a circular poloidal tokamak cross-section is shown on the left with
  computational perpendicular $s$-domains ($s=$ 2, 5, 8, 11, 14 of $n_s =16$
  total) mapped onto the 
  circle during the maximum growth phase ($t=38$ in units of $a/c_s$) of an
  ideal ballooning mode. On the right, the outer midplane area ($s=8$ with
  size $n_x=48$ and $n_y=512$) is
  magnified and the radial domain stretched by a factor of 2 for better
  visibility at times $t=38$ (max. flux) and $t=42$ (turbulent
  aftermath).}  
\end{figure}

The IBM ELM blowout scenario is similar to this, initially, except that the
instability is not a microinstability. Nevertheless, the scale differs by
less than an order of magnitude: the toroidal mode numbers for linear
ITG instabilities
are in the range of $n=30-100$ while the main ideal ballooning mode is near
mode number $n=10$, on the entire flux surface, for these typical
$\rho_s/L_{\perp}$ values (recall $\rho_s=\rho_i$ for $T_e=T_i$, and the
ratio $T_i/T_e$ is not far from unity in experimental cases).  
The ITG mode numbers are determined
by the ion
gyroradius and the dimensionless scale ratio $L_T/qR$, while the MHD
values are determined by the width of the pedestal.  Since $L_T$ is not
larger than about $30\rho_i$ in an H-mode pedestal, these scales are not
disparate. 

Due to the closeness of the native scales, the IBM instability very
quickly transfers energy to the ion gyroradius scale, which is only
about two or three steps away in a turbulence cascade which acts at a
factor of two at each step.  This brings the ITG and drift Alfv\'en
physics into play, and in any nonlinear stages the latter involves
stable shear Alfv\'en component with substantial energy content.
The moment of saturation is defined as the time at which the ion ExB
heat flux averaged over the computational domain is maximum; we refer to
this as the ``peak-flux'' time.  The results shown below indicate that
coincides with the establishment of the
fully turbulent regime.  Representation of this phase
requires resolution of all scales and inclusion of the
appropriate drift wave physics in the model.  How much resolution is
actually needed depends on the strength of the blowout, which is
situation dependent.  We therefore include a resolution scan on the
toroidal mode number spectrum.  The result is that 
convergence is reached in the temporal phases just after peak flux
only if the spectrum reaches down to $k_y\rho_i=1$.

The self-consistent equilibrium coupling in our gyrofluid model only
allows treatment of shifted circular $s-\alpha$ geometry (the difference
to MHD codes is that the high-resolution FLR-gyrofluid dynamics
necessitates field-aligned coordinates).  Effects of flux
surface shaping are postponed to later studies.  Our focus here is on
basic physics issues regarding the ability of the model to capture MHD
phenomena (see also Ref.\ \cite{Scott06} concerning global Alfv\`en
oscillations with self consistent profiles), resolution, turbulent
character, scalings and tests which are also accessible by $s-\alpha$
geometry. In particular, the theory of explosive instability has been
formulated for circular geometry \cite{Wilson04} and is therefore
accessible by the present model.

\section{Modelling of initial profiles}

First-principles based local drift wave edge turbulence simulations
are not able to obtain a realistic H-mode edge state with the known
experimental characteristics: correct density and temperature
pedestal profiles shapes or strength of flow shear are not obtained by
self-consistent evolution by specifying core sources only, nor has a threshold
transition character been found in any verified edge turbulence
simulation \cite{Scott07}. 
Therefore some kind of ``modelling'' has to take place when the IBM
instability (as an H-mode phenomenon) and its subsequent nonlinear evolution is
simulated with a nonlinear gyrofluid turbulence code: 
Although the realistic development of an edge transport barrier (and thus a
full ELM cycle) can not be directly obtained, one still may prescribe
the H-mode pedestal profile before the onset of an ELM, known from
experimental data, as an initial state for the simulation.  
As a base case for the prescribed pedestal profiles the well diagnosed edge
characteristics of ASDEX Upgrade H-mode shot \#17151 is used here
\cite{Horton05}. 

The local parameters, taken as mid pedestal values, correspond to electron and
ion temperatures $T_e = 300$~eV, $T_i = 360$~eV, densities $n_e = n_i = 2.5
\cdot 10^{19}$~m$^{-3}$, magnetic field strength $B=2.0$~T, major torus radius
$R = 1.65$~m, aspect ratio $R/a=3.3$, perpendicular temperature gradient
length $L_T = L_{\perp}=3.0$~cm, density gradient length $L_n = 6.0$~cm.
The profile of the
safety factor $q=1.5+3.5(r/a)^2$ is parabolic yielding local values at
the last closed flux surface (LCFS, $r/a=1.0$) of $q_a=5.0$ and 
$\hat s_a=1.4$.
The radial domain of the simulations covers a range of $L_{\perp}$
on either side of the LCFS.  The nominal pressure values are
$p_e=n_eT_e$ and $p_i=n_iT_i$.  All times are given in units of $a/c_s$.

The computational grid is given in terms of the $\{x,y,s\}$ coordinate
domains.  The spacing is equidistant.  For each grid point in the third
coordinate $s=s_k$ the $y$-coordinate is $y_k$ as defined in Eqs.\
(\ref{eqcoords},\ref{eqalphak}) so that for each grid point $s_k$ the
coordinate system $y=y_k$ is the one which has $g^{xy}_k=0$ at $s=s_k$.
For the nominal case $n_y=512$ perpendicular and $n_s=32$ parallel mesh points are
used. The radial domain with $n_x=64$ spans the plasma edge region between the
H-mode pedestal top, with plasma core parameters as inner boundary values, and
the 
outer bounded scrape-off layer region ($r/a=1\pm 0.06$). This represents a
spatial range from the global scale to smaller than the ion gyroradius
scale, with a ratio $\delta = \rho_s / a = 0.001875$ between ion gyroradius $
\rho_s$ and minor torus radius $a$.  In the SOL region ($r>a$) the
parallel boundary condition is replaced by a Debye sheath model whose
treatment is given in Ref.\ \cite{Ribeiro08}.

\begin{figure}
\includegraphics[width=7.5cm]{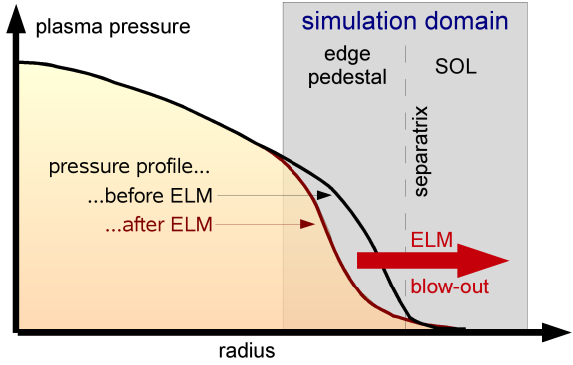}
\caption{\label{f:profiles} Radial tokamak edge pedestal profiles of the
  plasma pressure before and after an ELM blowout event across the separatrix
  into the scrape-off layer (SOL). The area shaded in grey shows the present
  radial numerical simulation domain.}
\end{figure}

\section{Profile pre-equilibration}

The initial conditions are thus prescribed and are based on experimentally
diagnosed radial
temperature and density pedestal profiles $T(r)$ and $n(r)$ for each species
(electrons and ions). A consistent electrostatic potential $\phi(r)$ is
derived by numerically solving the neoclassical balances (parallel
dynamics, toroidal drifts, and collisional dissipation, but not transport) in a
pre-processing step with a modified (zonally frozen) GEMR setup, resulting in
a time-steady 2D dissipative solution.  
The $s-\alpha$ MHD equilibrium in turn is set internally in GEMR by computing
the Shafranov shift, and in the present setup would not be consistently
described by
prescribed in terms of external or coupled shaped equilibrium solvers
\cite{Scott06}.

The parallel and perpendicular electron and ion temperatures, $T_e{}_\parallel$,
$T_e{}_\perp$, $T_i{}_\parallel$ and $T_i{}_\perp$, are directly adopted
and fixed from experimentally derived values by filtering the zonal component
out of the total time derivative $\partial T/\partial t = S_T$. 
Their values are zonally frozen by setting
\begin{equation}
{\partial T\over\partial t} = S_T - \langle S_T \rangle
\end{equation}
where $\langle S_T \rangle$ is the zonal (parallel and perpendicular) average
of all right-hand-side terms $S_T$ in the temperature evolution (Eqs.\ 5,6)
including numerical dissipation. 

On the other hand, gyrocenter 
densities $n_e$ and $n_i$ have to be set to obey relaxation
relations that allow the vorticity to freely evolve into equilibrium.
This is achieved by freezing the zonal component of the sum $n_e + \tau_i
n_i$, where $\tau_i = T_i / T_e$, as the density part of the pressure during
the equilibration phase, but allowing the difference (i.e., vorticity) to
evolve freely. 
This ensures that the contribution of densities to the total pressure is
zonally frozen through the relation 
\begin{equation}
{\partial\over\partial t} (n_e + \tau_i n_i) = S_e + \tau_i S_i - \langle S_e + \tau_i S_i
\rangle 
\end{equation}
while the densities relax regarding to
\begin{equation}
{\partial n_e\over\partial t} = S_e - {1 \over 1+\tau_i} \langle S_e + \tau_i S_i \rangle
\end{equation}
and
\begin{equation}
{\partial n_i\over\partial t} = S_i - {1 \over 1+\tau_i} \langle S_e + \tau_i S_i \rangle
\end{equation}

The numerical solution of the equilibration phase, starting directly from
realistically steep pedestal profiles $T_0(r)$, into steady state is delayed by
long, weakly damped global geodesic Alfv\'en oscillations. Convergence is
expedited by ramping up all of the gradients gradually from zero to prescribed
value over the first $\Delta t = \tau_r = 50 \; a /c_s$ of the run by
\begin{equation}
{\partial T\over\partial t} = S_T - \langle S_T \rangle + {1 \over \tau_r} T_0(x).
\end{equation}

This pre-processing equilibration phase is run until convergence 
with reduced perpendicular resolution ($n_x, n_y,
n_z$)=($64 \times 4 \times 32$), 
and without the ExB and magnetic flutter nonlinearities,
which allows establishment of the 2D
structure in a smooth manner. Then, the resolution is increased to the nominal
values ($64 \times 512 \times 32$), and a random turbulent bath with relative
amplitude $10^{-4} \; \rho_s / L_{\perp}$ is added to the background pedestal
profiles inside the closed flux surface region. This procedure reduces transient
Alfv\'enic and geodesic acoustic ringing and prepares a reproducible
initial state. However, the following sudden release of the nonlinearities also
leads to transient oscillations. Depending on parameters, these may still be
present at the onset of the instabilities, and can obscure a clean view
on the nonlinear growth rates, which will be relevant for the discussion below
on diagnosing linear or explosive instability.

A schematic sketch of the equilibrium pedestal profiles, representing an
idealised ASDEX Upgrade H-mode scenario \cite{Horton05}, is shown in
Fig.~\ref{f:profiles} together with the final relaxed state after the
ELM blowout phase (which is discussed in the next section).

\section{Computation of the ELM blowout}

When this initialised pedestal pressure profile is ideal ballooning unstable,
the IBM instability in GEMR simulations is observed to be
linearly growing in the pedestal region and at the onset of nonlinearity
further overshoots and saturates, representing the beginning of the
turbulent blowout phase during which a substantial fraction of the pedestal
energy is thrown onto the SOL.   Subsequent evolution involves
turbulence in both the pedestal and SOL regions as the original pedestal
energy is dissipated.

Previous nonlinear approaches on ELM ideal ballooning mode burst computations
have treated only the initial growth and nonlinear phases, focused on
low-wavelength modes and resolved only the MHD-relevant scales,
excluding treatment or discussion of the ion gyroradius scales.
This however precludes
the development of fully developed microturbulence which is caused by
the onset of nonlinearity (robust transfer of free energy to smaller
scales).   Ultimately, this microturbulence decides both the transition
to nonlinearity and the eventual saturation; that is, most of the
post-peak transport curve.  This affects the MHD scenario of
nonlinear explosive growth which most previous work advances.
However, we find that the exclusion of the ion gyroradius scales
produces an under-resolved situation.

We measure the quantitative character of the growth in both linear and
nonlinear phases with a growth curve 
\begin{equation}
\Gamma(t)={1\over 2Q_i} {\partial Q_i \over \partial t}
\end{equation}
defined in terms of
the heat flux $Q_i$ 
instead of the fluctuation free energy as explained above.
The heat flux is computed
as a zonal (flux surface) average,
\begin{equation}
Q_i(x) = \oint dy\,ds\LB 
(0.5\piplfl+\pippfl)\uex
+ (\pippfl+2n_i\tippfl)\wex\RB
\end{equation}
\begin{equation}
\uex=-{c F_0^{xy}\over B^2}\pyy{\phig} \qquad
\wex=-{c F_0^{xy}\over B^2}\pyy{\vorg}
\end{equation}
and then the time trace $Q_i(t)$, normalised in terms of $p_e c_s$ at
nominal parameters,
is computed as a volume average over
the central half of the radial domain.

The IBM instability, which follows the random seeding of the pre-processed
pedestal profile in our computations, is very violent, growing for the nominal
case at a rate $\Gamma = 0.18 c_s/L_{\perp}$,
just below the ideal interchange rate.  
On the other hand the growth curve 
$\Gamma(t)$ appears qualitatively the same as in basic turbulence cases.
At all time points in the nonlinear phase $\Gamma(t)$ is well below its
previous maximum.  Just after initial
saturation there is some nonlinear evolution
in which $\Gamma(t)$ crosses zero before settling down into long-term
statistical saturation.
At late times the initial blowout no longer imprints the
results: with a fixed source one merely finds bursty turbulence thereafter. 
Hence, there is no evidence for explosive instability in this nominal case,
which, if present, would be visible in a finite time singularity in the
fluctuating free energy $E \sim (\delta f)^2 \sim (t-t_{crit})^{-p} $ rising
with a power $p$. 

The
burst and the resulting decaying turbulence act to transport the
pedestal plasma across the LCFS, where energy and
particles are lost within short times by parallel boundary outflow to the
scrape-off layer (SOL) limiter. 
Without maintenance of a heat and particle source at the core boundary, 
this leads to decay of the initial gradient with a rapid onset at the
initial transition to nonlinearity at about $t=40$.  This time
point also corresponds to the peak-flux time.  

\begin{figure}
 \includegraphics[width=7.5cm]{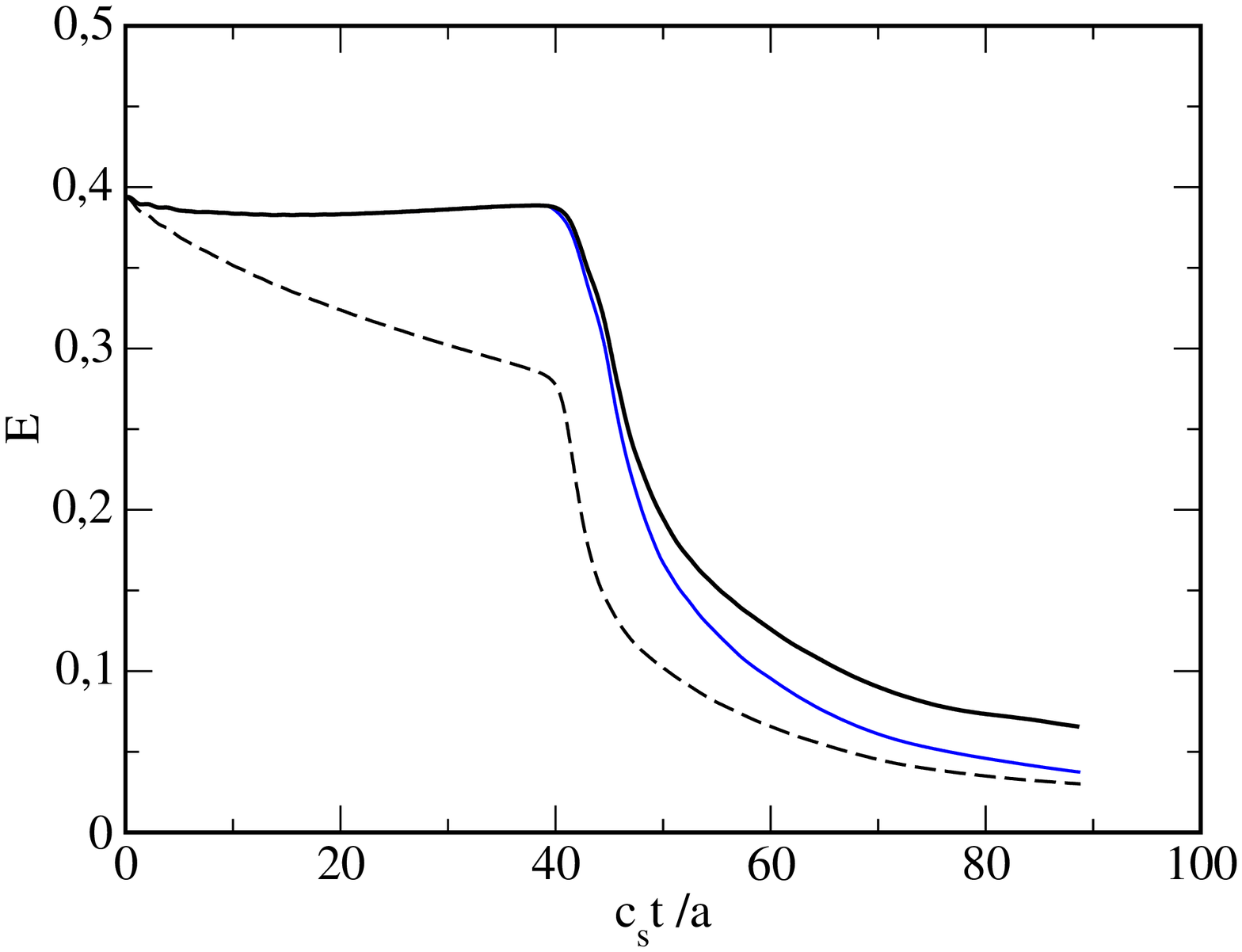}
 \includegraphics[width=7.5cm]{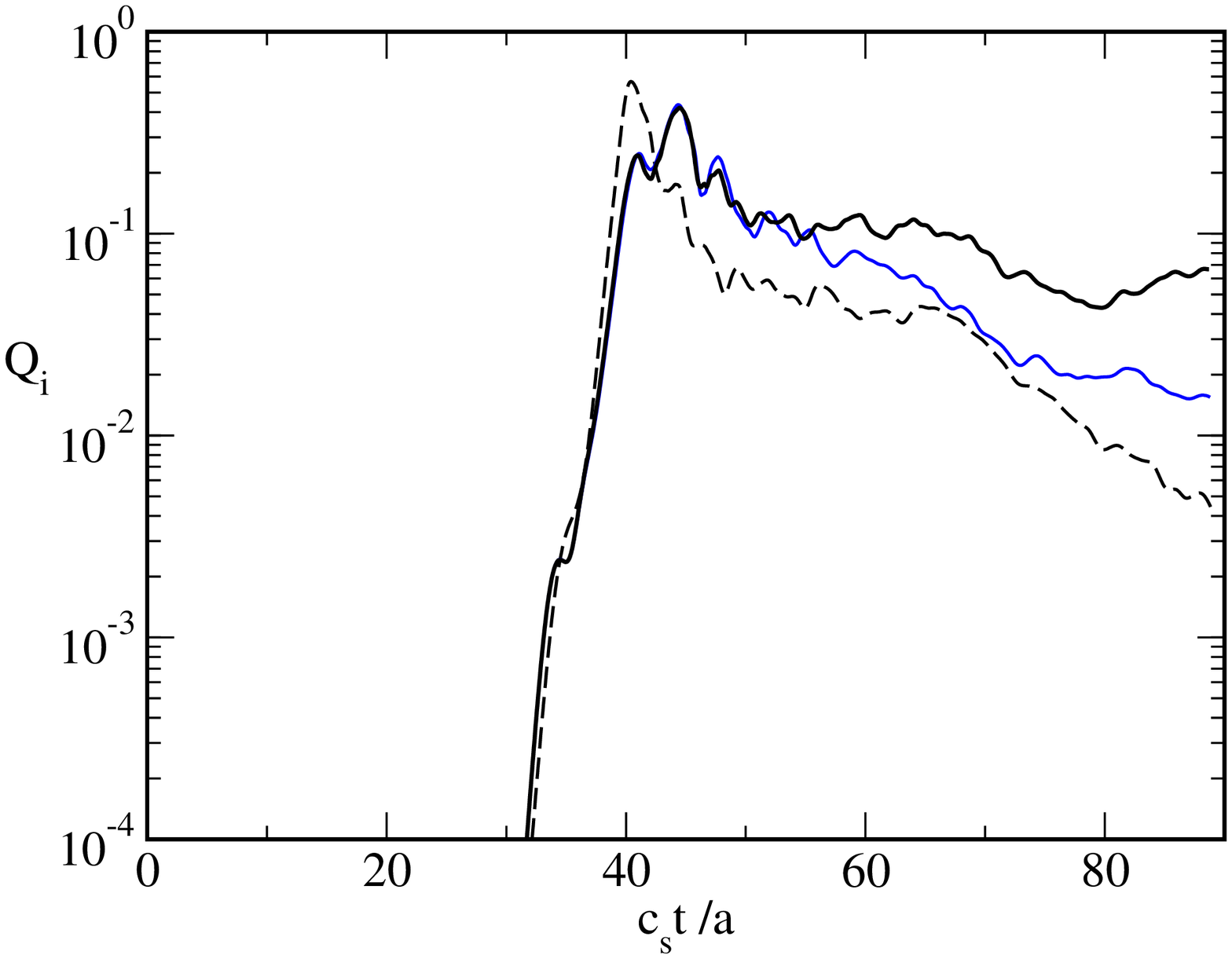}
\caption{\label{f:energy} Total energy $E(t)$ (top figure) and ion heat flux
  $Q_i(t)$ (bottom figure) during the ELM blowout: without source
  maintenance (dashed curve); with the source term switched off at the time of
  maximum instability (here: at $t = 40$); and with the source maintained
  for the whole simulation time.}
\end{figure}

However, the ion temperature profile already starts decaying during the
low-amplitude linear growth phase, due to neoclassical (finite Larmor
radius) transport 
(for detail on how this works in a fluid model see Ref.\
\cite{Falchetto07}). 
This situation is shown in Fig.~\ref{f:energy} (left frame, dashed line),
where the total pedestal energy is reduced by a quarter of its initial
content before the peak flux phase.   Therefore we test sensitivity
against sources.
Maintenance of the ion temperature profile by a fixed 
ion heat source localised to the inner boundary (following the
method for driven cases in Ref.\ \cite{Scott05c})
ensures nearly unchanged gradients
during the linear growth phase. 
At the end of the simulation time ($t=90$) 
after the ELM burst, the energy content is reduced to
around 10\% of the initial content when the source is absent or is switched off
at the peak flux time ($t=40$), and to around 20\% when the source is
maintained for the whole simulation time.

The influence of either switching off the source or maintaining the zonal
profiles on the nonlinear ion heat
flux $Q_i(t)$ is shown in  Fig.~\ref{f:energy}
(right): the linear growth phase and the peak flux is not significantly
changed by the source. As expected, the heat flux saturates for late
times with maintained core boundary inflow and is otherwise decaying.

\begin{figure}
 \includegraphics[width=7.5cm]{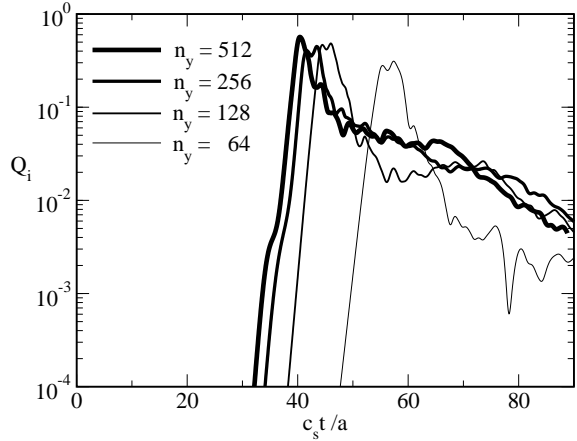}
\caption{\label{f:nyres} Consistency check for different perpendicular
  resolutions by varying the grid point number $n_y$ between 64 and
  512. Convergence is seen for $n_y=256$ and higher.}
\end{figure}

Consistency checks on the perpendicular resolution are made by
varying $n_y$ between 64, 128, 256 and 512. The evolution of $Q_i(t)$ for
these cases is shown in Fig.~\ref{f:nyres}. 
The $n_y=64$ case is clearly under-resolved and overestimates the total energy
decay by a factor 2.  For $n_y=128$ the initial part of the post-peak
decay phase disagrees with the $n_y=512$ case.  Convergence 
is found for $n_y=256$ and higher. The nonlinear saturation phase thus 
requires resolving the ion gyroradius scale.  This is the clearest
indication that turbulence for toroidal mode numbers beyond about
$30$ extending down to the $k_y\rho_i\sim 1$ range
is involved in the blowout saturation process.

\begin{figure}
 \includegraphics[width=7.5cm]{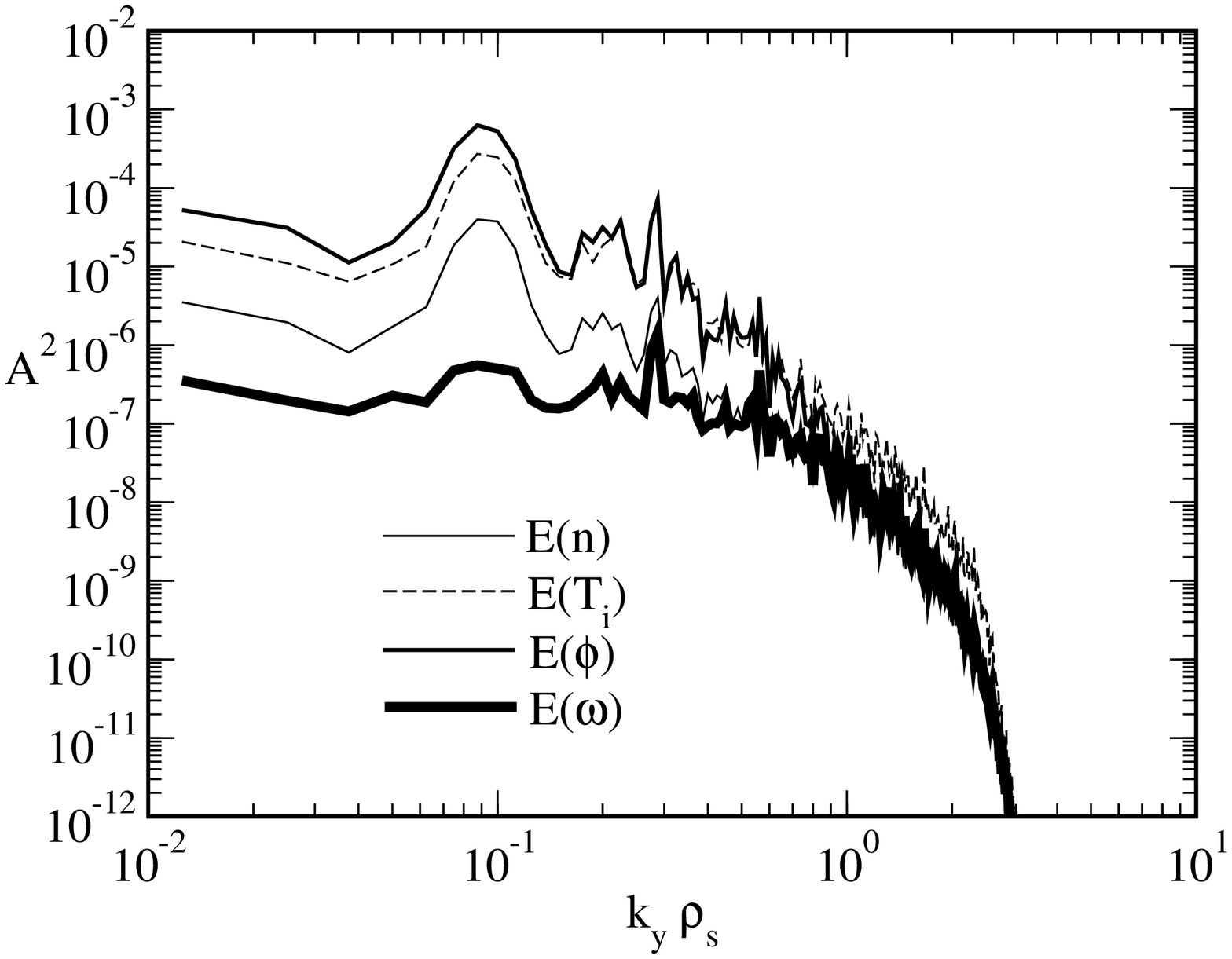}
 \includegraphics[width=7.5cm]{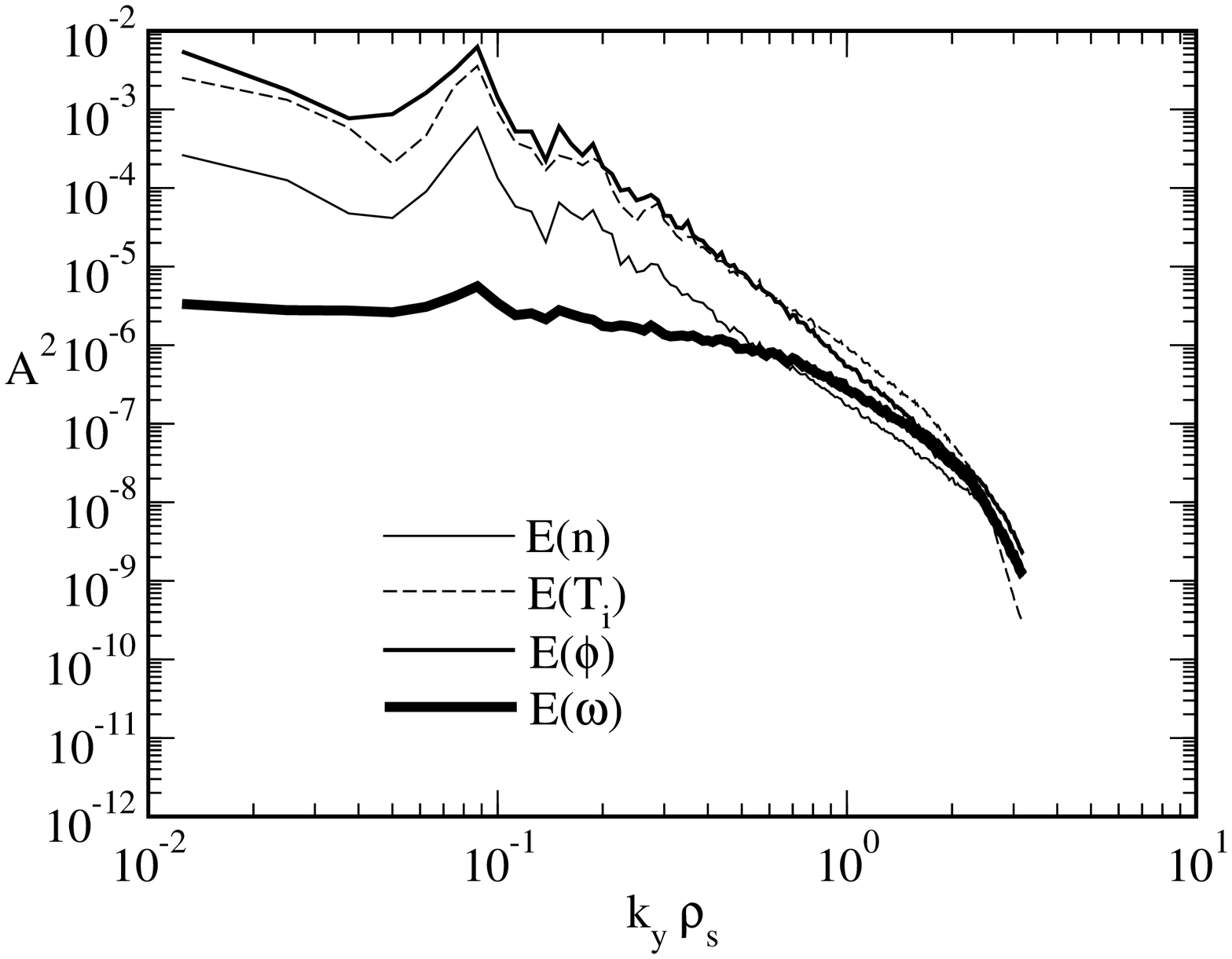}
\caption{\label{f:kyspecs} Perpendicular mode number $k_y \rho_s$ spectra for
  $t=38$ (top) and $t=42$ (bottom) around the peak flux phase. The vorticity
  spectrum (bold lines) is already flattening down to the ion gyroradius
  scale for $t=42$ only a few $\mu s$ after the maximum linear growth phase.} 
\end{figure}

The transition from linear instability to turbulence is studied during the peak
flux phase around $t=40$. Fig.~\ref{f:kyspecs} shows perpendicular mode number
$k_y \rho_s$ spectra of the squared amplitude of various fluctuating plasma
quantities (density $n$, ion temperature $T_i$, electrostatic potential $\phi$
and vorticity $\omega$) for $t=38$ and $t=42$. This time difference corresponds
to $\Delta t= 15 \mu s$ in physical units and around $70 L_{\perp}/c_s$ in
local drift units, which is only slightly faster than the overshoot and
saturation times known from edge microinstability cases \cite{Scott05b}.
Initially, the ion temperature gradient (ITG) driven microinstability and the
ideal ballooning mode (IBM) compete in growth out of the random low-amplitude
bath. The ITG mode grows strongest near the separatrix due to radially local
steepening by parallel SOL diffusion, which in our simulations may be
overestimated by using the standard fluid Bohm outflow boundary conditions. 
For $t = 38$ the linear IBM is clearly dominant near a toroidal mode number of
9-10 for the nominal AUG parameters, consistent with experimental observations
\cite{Kurzan05}. Around $t = 42$ rapid formation of a
turbulent cascade range in the spectra is observed and the vorticity spectrum is
already spread out to the ion gyroradius scale ($k_y \rho_s = 1$). 
This is a manifestation of the role of self
generated drift wave turbulence in the saturation process.  Because of
the way both the diffusive mixing and vorticity scattering
nonlinearities enter the drift wave physics \cite{Scott02} the spectrum
is held together as a unit,
all scales down to $\rho_i$ are involved in the saturation phase of the
overall ELM blowout transport.  The involvement of $\rho_i$ at unit
order, in turn, necessitates a nonlinear gyrofluid
model (and at some later date, a nonlinear gyrokinetic model).

\begin{figure}
\includegraphics[width=7.5cm]{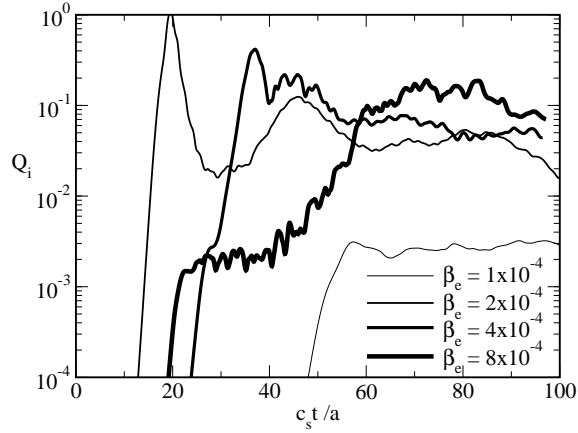}
\caption{\label{f:beta} Competing growth of ITG, microtearing and IBM
  instabilities determine onset times and intensities of the blowout
  time traces $Q_i(t)$ when scaling with the local plasma beta: for $\beta_e
  =10^{-4}$ only an ITG micro-instability is growing around $t=50$ and
  saturating on a low H-mode transport level. For $\beta_e = 2 \cdot 10^{-4}$ the
  IBM instability dominates and leads to a clean ELM signal around
  $t=20$. In the nominal case ($\beta_e=2 \cdot 10^{-4}$) ITG and IBM MHD growth
  compete initially with similar growth rates but the ideal ballooning mode
  takes over around $t=30$ and further determines the blowout. For $\beta_e=8
  \cdot 10^{-4}$ a microtearing instability is saturating on a gyro-Bohm
  transport level around $t=20$ and is transformed into more violent ITG MHD
  turbulence after deterioration of the initial electron temperature gradient
  ($t>50$) .
} 
\end{figure}

The characteristics of the blowout at various plasma beta is
studied for $0.5 \cdot 10^{-4}<\beta_e<8\cdot 10^{-4}$, with the
absolute $\beta$ a factor of $4.8$ higher due to the $T_i/T_e=1.2$ value.
A competition between unstable growth of ITG high-n modes and IBM low-n modes
is observed: For low $\beta$ an early onset of ITG turbulence degenerates the
profile and influences the strength of the IBM. For large $\beta$ the
nonlinear saturation is acting faster and stronger. The two effects lead to a
rather nondistinctive result concerning the peak ELM flux as shown in
Fig.~\ref{f:beta}, with a continuous transition between ITG and IBM triggered
onset of the turbulent aftermath.

The presence or absence of background current gradient terms, which in
this model are set by replacing
$(\tilde J_{||} \rightarrow \tilde J_{||} + J_0)$ everywhere the electron
parallel velocity $\tilde \nu_{||}$ appears in the equations, with $J_0$ given
by the $q$ profile, was found to have no discernible effect on the result.
Indeed self-consistent inclusion of $\grad J_0$ effects with possible
role in the L-to-H transition
was among the motivations of the GEMR model in the first place
\cite{Scott06}.  However, the magnitude of even an impressive pedestal
current peak is only one to three times the nominal saturation current
given by $n_e e c_s$, while values closer to $n_eec_sqR/L_{\perp}$ are
required to enter the energetics effectively.  Since $qR/L_{\perp}\sim
200$ for the nominal case, current gradient effects are very weak.

\section{Conclusions}

The main conclusion of this study is that the qualitative nature of the
saturation and aftermath of the initial IBM blowout is the same as for
generic edge turbulence given a small-amplitude start.  The transition
from linear mode structure and energetics to turbulence found for these
blowout cases is the same as in Ref.\ \cite{Scott05b}.
Only the nature
of the linear mode itself differs.  The blowout saturates upon its own
self generated drift-Alfv\'en turbulence, with a strong ion temperature
component given the gradients.  The vorticity spectrum reaches quickly
to the ion gyroradius ($\rho_i$)
scale, requiring the gyrofluid model and
explaining the numerical difficulties seen with Braginskii models --- it
can be argued that the cases given in Ref.\ \cite{Snyder05} 
crash on entry to the nonlinear stage.
Convergence in the aftermath requires resolving at least $\rho_i$.
Unfortunately, due to the lack of a self consistent H-mode state in a
well resolved computation, no threshold is found.  At lower beta values
one simply finds generic edge turbulence driven by the temperature
gradients.  It is not clear that this scenario really describes an
actual ELM, although the energetic growth and decay curves are not
unrealistic.  However, with a well resolved transition to nonlinearity
in both the energetic peak and aftermath phases, our studies find no
evidence of an explosive MHD phenomenology.  Indeed it can be argued
that nonlinear MHD processes are pre-empted by the efficient transfer to
smaller scales through the two-fluid drift wave physics more commonly
associated with microturbulence.

\section*{Acknowledgements}
This work was supported by the Austrian Science Fund FWF under contract
Y398, by a junior research grant (``Nachwuchsf\"orderung'') from University of
Innsbruck, and by the European Communities under the Contract of
Associations between Euratom and the Austrian Academy of Sciences, 
carried out within the framework of the European Fusion Development
Agreement. The views and opinions herein do not necessarily reflect those of
the European Commission.



\begin{thebibliography}{10}

\bibitem{Zohm96}
H. Zohm,
\newblock Plasma Phys. Control. Fusion {\bf 38}, 105 (1996).

\bibitem{Connor98a}
J.W. Connor,
\newblock Plasma Phys. Control. Fusion {\bf 40}, 531 (1998).

\bibitem{Becoulet03}
M. Becoulet, G. Huysmans, Y. Sarazin et al.,
\newblock Plasma Phys. Control. Fusion {\bf 45}, A93 (2003).

\bibitem{ITER}
K. Ikeda et al.,
\newblock Nucl. Fusion {\bf 47}, E01 (2007).

\bibitem{Lang03}
P.T. Lang, J. Neuhauser, L.D. Horton et al.,
\newblock  Nucl. Fusion {\bf 43}, 1110 (2003).

\bibitem{Evans06}
T.E. Evans, R.A. Moyer, K.H. Burrell et al.,
\newblock Nature Physics {\bf 2}, 419 (2006).

\bibitem{Hand08}
E. Hand,
\newblock Nature {\bf 452}, 11 (2008).

\bibitem{Wagner82}
F. Wagner, G. Becker, K. Behringer et al.,
\newblock Phys. Rev. Lett. {\bf 49}, 1408 (1982). 

\bibitem{Kamiya07}
K. Kamiya, N. Asakura, J. Boedo et al.,
\newblock Plasma Phys. Control. Fusion {\bf 49}, S43 (2007).

\bibitem{Wagner06}
V. Erckmann, F. Wagner, J. Baldzuhn et al.,
\newblock Phys. Rev. Lett. {\bf 70}, 2086 (1993). 

\bibitem{Eich03}
T. Eich, A. Herrmann, J. Neuhauser et al.,
\newblock Phys. Rev. Lett. {\bf 91}, 195003 (2003). 

\bibitem{Eich07}
T. Eich, A. Kallenbach, R.A. Pitts et al.,
\newblock J. Nucl. Materials {\bf 363-365}, 989 (2007). 

\bibitem{Eich09}
T. Eich, A. Kallenbach, W. Fundamenski et al.,
\newblock J. Nucl. Materials {\bf 390-391}, 760 (2009). 

\bibitem{Wilson04}
H.R. Wilson and S.C. Cowley,
\newblock Phys. Rev. Lett. {\bf 92}, 175006 (2004). 

\bibitem{Fundamenski07}
W. Fundamenski, V. Naulin, T. Neukirch et al.,
\newblock Plasma Phys. Control. Fusion {\bf 49}, R43 (2007).

\bibitem{Bath85}
G.T. Bath, 
\newblock Rep. Prog. Phys {\bf 48}, 483 (1985).

\bibitem{Connor98b}
J.W. Connor,
\newblock Plasma Phys. Control. Fusion {\bf 40}, 191 (1998).

\bibitem{Scott06}
B.D. Scott,
\newblock  Contrib. Plasma Phys. {\bf 46}, 714 (2006).

\bibitem{Scott07}
B.D. Scott,
\newblock  Plasma Phys. Control. Fusion {\bf 49}, S25 (2007).

\bibitem{Snyder02}
P. Snyder, H. Wilson, J. Ferron, et al.
\newblock Phys. Plasmas {\bf 9}, 2037 (2002).

\bibitem{Wilson06}
H. Wilson, S. Cowley, A. Kirk, P. Snyder
\newblock Plasma Phys. Control. Fusion {\bf 48} A71 (2006).

\bibitem{Snyder07}
P.B. Snyder, K.H. Burrell, H.R.Wilson, et al.
\newblock Nucl. Fusion {\bf 47}, 961 (2007).

\bibitem{Snyder05}
P. Snyder, H. Wilson, X. Xu
\newblock Phys. Plasmas {\bf 12}, 056115 (2005)

\bibitem{Brennan06}
D.P. Brennan, S.E. Kruger, D.D. Schnack, C.R. Sovinec, A. Pankin,
\newblock J. Phys. Conf. Series {\bf 46}, 63 (2006).

\bibitem{Strauss06}
H. Strauss, L. Sugiyama L., C.S. Chang, et al.
\newblock Proc. 21st IAEA Fusion Energy Conf. (Chengdu, China, 2006) (Vienna:
IAEA) CD-ROM file TH/P8-6 and
http://www-naweb.iaea.org/napc/physics/FEC/FEC2006/html/index.htm 

\bibitem{Pankin07}
A.Y. Pankin, G. Bateman, D.P. Brennan, A.H. Kritz, S. Kruger,
P.B. Snyder, C. Sovinec and the NIMROD team,
\newblock Plasma Phys. Control. Fusion {\bf 49}, S63 (2007).

\bibitem{Mizuguchi07}
N. Mizuguchi, R. Khan, T. Hayashi and N. Nakajima
\newblock Nucl. Fusion {\bf 47}, 579 (2007)

\bibitem{Huysmans07}
G.T.A. Huysmans and O. Czarny,
\newblock Nucl. Fusion {\bf 47}, 659 (2007). 

\bibitem{Scott05}
B.D. Scott,
\newblock Phys. Plasmas {\bf 12}, 102307 (2005).

\bibitem{Scott07b}
B.D. Scott,
\newblock Phys. Plasmas {\bf 14}, 102318 (2007).

\bibitem{Strauss76}
H. Strauss,
\newblock Phys. Fluids {\bf 19}, 134 (1976).

\bibitem{Zweben09}
S. Zweben, B. Scott, J. Terry et al.,
\newblock Phys. Plasmas {\bf 16}, 082505 (2009).

\bibitem{Dorland93}
W. Dorland and G. Hammett,
\newblock Phys. Fluids B {\bf 5}, 812 (1993).

\bibitem{Beer96}
M. A. Beer and G. Hammett,
\newblock Phys. Plasmas {\bf 3}, 4046 (1996).

\bibitem{gem2}
B.D. Scott,
Derivation via free energy conservation constraints of gyrofluid equations
with finite-gyroradius electromagnetic nonlinearities,
\newblock submitted to Phys. Plasmas, arXiv:0710.4899 (2007).

\bibitem{Braginskii}
S.I. Braginskii, 
\newblock Rev. Plasma Phys. {\bf 1}, 205 (1965).


\bibitem{Kendl06}
A. Kendl and B.D. Scott,
\newblock Phys. Plasmas {\bf 13}, 012504 (2006).

\bibitem{Scott01}
B.D. Scott,
\newblock  Phys. Plasmas {\bf 8}, 447 (2001).

\bibitem{Scott02}
B.D. Scott,
\newblock New Journal of Physics {\bf 4}, 52 (2002).

\bibitem{Huysmans09}
G.T.A. Huysmans,
\newblock submitted to Plasma Phys. Control. Fusion {\bf 51}, (2009).

\bibitem{Scott05b}
B.D. Scott,
\newblock Phys. Plasmas {\bf 12}, 062314 (2005).

\bibitem{Horton05}
L.D. Horton, A.V. Chankin, Y.P. Chen et al.,
\newblock Nucl. Fusion {\bf 45}, 856 (2005).

\bibitem{Falchetto07}
G. L. Falchetto, M. Ottaviani, X. Garbet, and A. Smolyakov,
\newblock  Phys. Plasmas {\bf 14}, 082304 (2007).

\bibitem{Ribeiro08}
T. Ribeiro and B. Scott,
\newblock Plasma Phys. Control. Fusion {\bf 50}, 055007 (2008).

\bibitem{Scott05c}
B.D. Scott,
\newblock Phys. Plasmas {\bf 12}, 082305 (2005).

\bibitem{Kurzan05}
B. Kurzan, H.D. Murmann, J. Neuhauser,
\newblock Phys. Rev. Lett. {\bf 95}, 145001 (2005). 


\end{thebibliography}

\end{document}